\documentclass[11pt]{article}
\usepackage[left=1in,top=1in,right=1in,bottom=1in,head=.1in,nofoot]{geometry}
\usepackage{natbib,setspace,amsmath,bm,graphicx,algpseudocode,enumitem,appendix}
\graphicspath{ {Graphics/} }
\setlist[itemize,enumerate]{leftmargin=*} 
\usepackage{etoolbox} 
\apptocmd{\sloppy}{\hbadness 10000\relax}{}{} 

\begin{document}
\title{More Powerful Multiple Testing in Randomized Experiments with Non-Compliance}
\author{Joseph J. Lee, Laura Forastiere, Luke Miratrix, and Natesh S. Pillai \\ Department of Statistics, Harvard University \\ Cambridge, MA 02138}
\maketitle

\begin{abstract}
Two common concerns raised in analyses of randomized experiments are (i) appropriately handling issues of non-compliance, and (ii) appropriately adjusting for multiple tests (e.g., on multiple outcomes or subgroups).  Although simple intention-to-treat (ITT) and Bonferroni methods are valid in terms of type I error, they can each lead to a substantial loss of power; when employing both simultaneously, the total loss may be severe.  Alternatives exist to address each concern.  Here we propose an analysis method for experiments involving both features that merges posterior predictive $p$-values for complier causal effects with randomization-based multiple comparisons adjustments; the results are valid familywise tests that are doubly advantageous: more powerful than both those based on standard ITT statistics and those using traditional multiple comparison adjustments.  The operating characteristics and advantages of our method are demonstrated through a series of simulated experiments and an analysis of the United States Job Training Partnership Act (JTPA) Study, where our methods lead to different conclusions regarding the significance of estimated JTPA effects.
\end{abstract}

\section{Introduction}
The United States Job Training Partnership Act (JTPA) Study was a randomized experiment in the 1980s designed to measure the effects of a national, publicly-funded training program.  Participants randomly assigned to the treatment group were eligible to receive JTPA services, while participants randomly assigned to the control group were barred from JTPA services for 18 months.  Only about 2/3 of the treatment participants, however, actually enrolled and received any JTPA services; the other 1/3 failed to comply with their treatment assignment.  Furthermore, because of the fluid nature of the participants' employment, researchers were interested in measuring JTPA effects across several time periods after random assignment, including the in-training period and the first and second post-program years.  Analyzing such data requires addressing two substantial concerns: (i) due to non-compliance, the effects of treatment assignment are not equivalent to the effects of treatment receipt, and (ii) conducting tests for multiple time periods without appropriate adjustments may lead to an inflated type I error rate.  In this paper, we outline an analysis method that addresses both concerns while maintaining reasonable power to detect treatment effects.

When units in randomized experiments fail to comply with their random assignment, inference for the effects of treatment receipt, rather than of assignment alone, becomes less straightforward.  Intention-to-treat (ITT) analyses, which ignore treatment receipt, may have low power when assignment alone has no effect on the experimental outcome.  In order to address this loss of power, \citet*{Rubin1998} introduced randomization-based posterior predictive $p$-values for the complier average causal effect (CACE) and showed through simulation that (i) they are valid $p$-values in terms of type I error, and (ii) their tests have higher power than tests using ITT $p$-values under reasonable alternative hypotheses (e.g., hypotheses with non-zero treatment effects for units who are assigned to and receive treatment, but zero treatment effects for units who do not receive it).  This framework follows the general approach for Bayesian causal inference in randomized experiments with non-compliance outlined by \citet*{Imbens1997}.  Both pieces of work rely on the multiple imputation \citep*{Rubin1987} of missing compliance statuses; separating the experimental units into principal strata \citep*{Frangakis2002} based on compliance behavior aids inference for the desired causal effect.  We use these tools in our approach but adapt them for simultaneous testing of multiple outcomes and subgroups.

Multiple testing issues are common in randomized experiments because multiple outcomes and subgroups of interest are often measured and analyzed for possible effects.  Traditionally, practitioners have applied Bonferroni corrections to sets of $p$-values in order to control their familywise error rate (FWER), i.e., the rate at which at least one type I error is made, in a straightforward manner.  Bonferroni corrections, however, tend to be overly conservative, especially when those $p$-values are correlated \citep*{Westfall1989}.  This fact has led many applied researchers to avoid Bonferroni corrections and abandon multiple comparisons adjustments altogether \citep*{Cabin2000,Nakagawa2004,Perneger1998,Rothman1990}.  Other avenues exist; randomization-based procedures can provide greater power while maintaining the FWER by accounting for correlated tests.  \citet*{Brown1981} and \citet*{Westfall1989} first introduced permutation-based multiple testing adjustments, though they did not explicitly motivate them using randomized assignment mechanisms.  Randomization-based procedures are additionally appealing because they do not require any assumptions about the underlying distribution (here, joint) of the data.  Furthermore, recent increases in computational power have helped such procedures become more tractable and gain popularity \citep*{Good2005}.

In this article, we connect methodological ideas to appropriately handle both non-compliance and multiple testing in randomized experiments.  We build up to this combined approach in stages.  In Section \ref{sec:Non-compliance}, we elucidate the method proposed by \citet*{Rubin1998} for evaluating meaningful causal effects in the presence of non-compliance.  In Section \ref{sec:Multiple}, we extend the ideas of \citet*{Westfall1989} to fully randomization-based multiple comparisons adjustments and propose such adjustments as a straightforward yet more powerful alternative to Bonferroni corrections.  In Section \ref{sec:combined}, we merge the notions of non-compliance and multiple testing, and outline a combined method of analysis that demonstrates power advantages from both perspectives.  In each of Sections \ref{sec:Non-compliance}--\ref{sec:combined}, we empirically show the benefits of the described methods through a series of simulated experiments.  In Section \ref{sec:JTPA}, we apply traditional methods and our combined method to JTPA data to evaluate the program's effects on employment rate by time period.  We illustrate how the methods lead to different conclusions regarding the significance of estimated JTPA effects.  Section \ref{sec:conclusion} concludes.

\section{Experiments with Non-compliance}\label{sec:Non-compliance}
\subsection{Non-compliance as a missing data problem}
Suppose we have a randomized experiment with $N$ units, indexed by $i$, with observed covariates $X_i$, randomly assigned to control or active treatment.  Let $Z_i$ be a binary indicator for assignment to active treatment, and let $D_i(z)$ be a binary indicator for receipt of active treatment under assignment $z$.  A unit's compliance behavior $C_i$ is defined by the pair of potential outcomes \citep*{Neyman1923,Rubin1974} $(D_i(0),D_i(1))$; this notation is adequate under the stable unit treatment value assumption \citep*{Rubin1980,Rubin1986}, which asserts no interference between experimental units, as well as two well-defined outcomes.  Each unit then belongs to one of four possible compliance strata:
\begin{itemize}
\item Compliers ($C_i=c$), who receive their treatment assignment: $(D_i(0),D_i(1))=(0,1)$.
\item Never-takers ($C_i=nt$), who never receive the active treatment: $(D_i(0),D_i(1))=(0,0)$.
\item Always-takers ($C_i=at$), who always receive the active treatment: $(D_i(0),D_i(1))=(1,1)$.
\item Defiers ($C_i=d$), who receive the opposite of their treatment assignment: $(D_i(0),D_i(1))=(1,0)$.
\end{itemize}

If non-compliance is one-sided --- i.e., units assigned to control are prohibited from receiving the active treatment --- then $D_i(0)=0$ for all $i$.  In such settings, always-takers and defiers do not exist, and two possible strata are left: compliers and never-takers.  Real-world scenarios involving one-sided non-compliance include many clinical trials, in which new drugs are unavailable to control patients, and some job training experiments, in which training programs and additional services are unavailable to the control group.

In many practical settings, researchers are most interested in the compliers because the effect of treatment assignment is synonymous with the effect of treatment receipt for those units.  Strata membership, however, can never be fully determined for all units because they depend on the two potential outcomes of $D$, one of which is missing (i.e., unobserved).  Membership can, on the other hand, be partially determined based on the observed potential outcome, $D_i^\mathrm{obs}$.  Table \ref{tbl:ComplianceStrata} outlines the possible compliance strata based on units' observed treatment assignment and receipt.  An example ``Science" table \citep*{Rubin2005} under one-sided non-compliance and its observed values under a particular assignment are shown in Table \ref{tbl:science}.

\begin{table}[!ht]
\centering
\begin{tabular}{c c c c c}
  \hline
  Assignment & Receipt & \multicolumn{2}{c}{Possible $C_i$ Values} \\ \hline
  $Z_i$ & $D_i^{\mathrm{obs}}$ & One-sided Non-compliance & Two-sided Non-compliance \\ \hline \hline
  0 & 0 & $c, nt$ & $c,nt$ \\
  0 & 1 & -- & $at, d$ \\
  1 & 0 & $nt$ & $nt, d$ \\
  1 & 1 & $c$ & $c, at$ \\
  \hline  
\end{tabular}
\caption{Units' possible compliance strata based on observed treatment assignment and receipt.}
\label{tbl:ComplianceStrata}
\end{table}

\begin{table}[!ht]
\centering
\resizebox{16cm}{!}{
\begin{tabular}{c c c c c c c | c c c c c c}
  \hline
  & & \multicolumn{2}{c}{$D(z)$} & Compliance & \multicolumn{2}{c |}{$Y(z)$} & Assignment & \multicolumn{2}{c}{$D(z)$} & Compliance & \multicolumn{2}{c}{$Y(z)$} \\ \cline{3-13}
  Unit & $X_i$ & $D_i(0)$ & $D_i(1)$ & $C_i$ & $Y(0)$ & $Y(1)$ & $Z_i$ & $D_i(0)$ & $D_i(1)$ & $C_i$ & $Y_i(0)$ & $Y_i(1)$ \\ \hline \hline
  1 & $X_1$ & 0 & 0 & $nt$ & $Y_1(0)$ & $Y_1(1)$ & 0 & 0 & ? & ? & $Y_1^{\mathrm{obs}}$ & ? \\
  2 & $X_2$ & 0 & 1 & $c$ & $Y_2(0)$ & $Y_2(1)$ & 1 & 0 & 1 & $c$ & ? & $Y_2^{\mathrm{obs}}$ \\
  3 & $X_3$ & 0 & 1 & $c$ & $Y_3(0)$ & $Y_3(1)$ & 1 & 0 & 1 & $c$ & ? & $Y_3^{\mathrm{obs}}$ \\
  4 & $X_4$ & 0 & 0 & $nt$ & $Y_4(0)$ & $Y_4(1)$ & 1 & 0 & 0 & $nt$ & ? & $Y_4^{\mathrm{obs}}$ \\
  \ldots & & \multicolumn{5}{c |}{\ldots} & \multicolumn{6}{c}{\ldots} \\
  $N$ & $X_N$ & 0 & 1 & $c$ & $Y_N(0)$ & $Y_N(1)$ & 0 & 0 & ? & ? & $Y_N^{\mathrm{obs}}$ & ? \\
  \hline  
\end{tabular}
}
\caption{An example Science table under one-sided non-compliance (left) and its corresponding observed and unobserved values under a particular assignment (right).}
\label{tbl:science}
\end{table}

Because strata memberships are not fully observed, uncertainty with respect to complier-specific effects stems from the missing compliance statuses (i.e., $D$ potential outcomes) in addition to the missing $Y$ potential outcomes.  One approach to handling the additional uncertainty is to, in a Bayesian framework, view the missing compliance statuses as random variables.  By multiply imputing the missing compliance statuses, e.g., according to a distributional model, they can be integrated out, and we can make inference specific to the compliers.

\subsection{Randomization-based posterior predictive $p$-values}\label{sec:ppp}
As described by \citet*{Meng1994}, a posterior predictive $p$-value can be viewed as the posterior mean of a classical $p$-value, averaging over the posterior distribution of nuisance factors (e.g., missing compliance statuses) under the null hypothesis.  \citet*{Rubin1998} introduced a randomization-based procedure, which we expound on here, for obtaining posterior predictive $p$-values for estimated complier-only effects.  One posterior predictive $p$-value is the average of many $p$-values calculated from multiple ``compliance-complete" datasets with imputed compliance statuses; for each compliance-complete dataset, the $p$-value is obtained through a randomization test \citep*{Fisher1925,Fisher1935}.

Within one randomization test, however, calculations of the test statistic do not use all of the compliance information from the compliance-complete data; rather, they use only the compliance information that would have actually been observed under particular hypothetical randomizations.  Though implied, this step of re-observing the data is not explicitly stated by \citet*{Rubin1998}; we place it in Step 5 of the procedure below for emphasis because it is an important prerequisite for conducting a proper test.  Unlike discrepancy variables \citep*{Meng1994}, which may depend on unobserved factors (e.g., missing compliance statuses), test statistics must be functions of only the observed data.  In order to conduct a proper test, the true observed test statistic value must be measured against the correct distribution, i.e., the distribution of that same test statistic.

In this section, we assume a single outcome for simplicity.  The procedure for obtaining a randomization-based posterior predictive $p$-value is as follows:
\begin{enumerate}
\item \textbf{Choose a test statistic and calculate its observed value.} \newline Choose a test statistic, $T$, to estimate an effect on the outcome, $Y$.  Examples include the maximum-likelihood estimate (MLE) of CACE or the posterior median of CACE, given the observed compliance statuses and potential outcomes, under the exclusion restriction \citep*[see][]{Angrist1996,Imbens1997}.  Calculate $T$ on the observed data to obtain $T^\mathrm{obs}$.
\end{enumerate}
\begin{algorithmic}
\For{$m:1$ to $M$}
\begin{enumerate} \setcounter{enumi}{1}
\item \textbf{Impute missing compliance statuses.} \newline Impute the missing compliance statuses, drawing once from their posterior predictive distribution according to a compliance model that assumes the null hypothesis (e.g., of zero effect).
\item \textbf{Impute missing potential outcomes.} \newline Impute the missing $Y$ potential outcomes under the sharp null hypothesis.  Under the typical sharp null hypothesis of zero treatment effect, the missing potential outcome for unit $i$ is imputed exactly as $Y_i^\mathrm{obs}$.
\item \textbf{Draw a random hypothetical assignment.} \newline Draw a random hypothetical assignment vector according to the assignment mechanism used in the original experiment.
\item \textbf{Re-observe the data.} \newline Treating the imputed compliance statuses, imputed potential outcomes, and hypothetical assignment vector from Steps 2--4 as true, create a corresponding hypothetical observed dataset by masking the potential outcomes and compliance statuses that would not have been observed under the hypothetical assignment.
\item \textbf{Calculate the test statistic on these data.} \newline Calculate $T$ on the hypothetical observed data to obtain $T^{\mathrm{hyp}}$.  Record whether this statistic is at least as extreme as $T^{\mathrm{obs}}$.
\end{enumerate}
\EndFor
\end{algorithmic}
\begin{enumerate} \setcounter{enumi}{6}
\item \textbf{Calculate the posterior predictive $p$-value.} \newline The posterior predictive $p$-value for the null hypothesis with respect to $T$ equals the proportion of the $M$ imputation-randomization sets for which $T^{\mathrm{hyp}}$ is as extreme as or more extreme than $T^{\mathrm{obs}}$.
\end{enumerate}

\citet*{Rubin1998} discusses several commonly used statistics for evaluating complier causal effects, only some of which tend to estimate the CACE and thus provide suitable power against appropriate alternative hypotheses.  As is commonly done in the non-compliance literature, we assume the exclusion restriction (i.e., we assume that treatment assignment has no effect on the outcomes of never-takers and always-takers) for test statistic calculations throughout this paper.  Such an assumption is not necessary and does not affect the validity of the randomization test, but it does facilitate more precise estimation of CACE when true \citep*[see][]{Imbens1997} and is often reasonable.

The imputation in Step 2 is performed probabilistically, using the missing statuses' null posterior predictive distribution, given $X,Z,D^\mathrm{obs},$ and $Y^\mathrm{obs}$.  (Some test statistics, such as the posterior median of CACE, may be computed by multiply imputing the missing compliance statuses.  This would be a separate imputation from the one described in Step 2 above.  If the test statistic calculation itself involves imputation, such imputation does not need to, and usually does not, assume the null hypothesis.)  The repetition of Steps 2--6 is intended to reflect the uncertainty of estimation resulting from the missing compliance statuses; $M$ is a large number (e.g., $10,000$) that controls the Monte Carlo integration error.

Under the null hypothesis, $Y$ is not affected by assignment to or receipt of the active treatment; it is therefore treated like a covariate in the imputation model.  Even in the absence of other covariates ($X$), $Y$ alone may still be successful in stochastically identifying the missing compliance statuses, thus providing tests of CACE with power over ITT tests (see Section \ref{sec:simNoncompliance}).  When additional covariates that affect compliance status supplement $Y$ in the imputation model (e.g., in a Bayesian generalized linear model), the compliance identification tends to sharpen, providing CACE tests with greater power.

In settings with one-sided non-compliance, only the compliance statuses of units assigned to the control group are missing.  Let $\omega_c$ be the super-population proportion of compliers, and let $\bm{\eta}=(\eta_c,\eta_n)$ be the parameters that govern the outcome distributions of compliers and never-takers, respectively.  Note that under the null hypothesis, these are only two outcome distributions; units within a compliance stratum have the same outcome distributions, regardless of their treatment assignment.  The posterior predictive distribution of the missing compliance statuses can be obtained using a two-step data augmentation algorithm \citep*{Tanner1987}.  Using the current (or initial, if starting the algorithm) values of the parameters, the missing compliance statuses are drawn according to Bayes' rule:
\begin{equation}\label{eqn:compprob}
P(C_i=c|Y_i^{\mathrm{obs}},X_i,Z_i=0,D_i^\mathrm{obs}=0,\omega_c,\bm{\eta})=
\frac{\omega_c g_c(Y_i^{\mathrm{obs}}; \eta_c)}{\omega_c g_c(Y_i^{\mathrm{obs}}; \eta_c)+(1-\omega_c) g_n(Y_i^{\mathrm{obs}}; \eta_n)},
\end{equation}
where $g_c(y; \eta_c)$ and $g_n(y; \eta_n)$ are the outcome probabilities (or densities) of $y$ for compliers and never-takers, respectively.  Once the missing compliance statuses are drawn, new parameter values are drawn from their compliance-complete-data posterior distributions.  These two steps are alternated until distributional convergence.  After convergence, the draws of the missing compliance statuses can be treated as posterior predictive imputations.  Obtaining posterior draws of parameters --- and consequently, posterior predictive draws of the missing compliance statuses --- may be more straightforward if models are conjugate, e.g., Beta-Binomial or Dirichlet-Multinomial models (see Section \ref{sec:simNoncompliance}).

For each imputation of the missing compliance statuses, a randomization test (here involving only one random hypothetical assignment for computational efficiency) is performed in Steps 3--6.  Because $p$-values are defined as conditional probabilities given that the sharp null hypothesis is true, the imputation of $Y$ potential outcomes in Step 3 must occur under this hypothesis.  Table \ref{tbl:obsScienceNull} shows the observed values of the Science table from Table \ref{tbl:science}, with the $Y$ potential outcomes imputed under the sharp null hypothesis of zero treatment effect.  For computational efficiency, Step 3 can be performed just once (before the loop) because this imputation is deterministic.

\begin{table}[!ht]
\centering
\begin{tabular}{c c c c c c c c}
  \hline
  & & Assignment & \multicolumn{2}{c}{$D(z)$} & Compliance status & \multicolumn{2}{c}{$Y(z)$} \\ \cline{3-8}
  Unit & $X_i$ & $Z_i$ & $D_i(0)$ & $D_i(1)$ & $C_i$ & $Y_i(0)$ & $Y_i(1)$ \\ \hline \hline
  1 & $X_1$ & 0 & 0 & ? & ? & $Y_1^\mathrm{obs}$ & $(Y_1^\mathrm{obs})$ \\
  2 & $X_2$ & 1 & 0 & 1 & $c$ & $(Y_2^\mathrm{obs})$ & $Y_2^\mathrm{obs}$ \\
  3 & $X_3$ & 1 & 0 & 1 & $c$ & $(Y_3^\mathrm{obs})$ & $Y_3^\mathrm{obs}$ \\
  4 & $X_4$ & 1 & 0 & 0 & $nt$ & $(Y_4^\mathrm{obs})$ & $Y_4^\mathrm{obs}$ \\
  \ldots & & & & & \ldots & & \\
  $N$ & $X_N$ & 0 & 0 & ? & ? & $Y_N^\mathrm{obs}$ & $(Y_N^\mathrm{obs})$ \\
  \hline  
\end{tabular}
\caption{The observed values of the Science table from Table \ref{tbl:science}, with the missing $Y$ potential outcomes imputed under the sharp null hypothesis of zero treatment effect.  Imputed $Y$ potential outcomes are in parentheses.}
\label{tbl:obsScienceNull}
\end{table}

The random draw of a hypothetical assignment vector in Step 4 depends on the specific assignment mechanism used in the experiment, e.g., complete randomization or block randomization.  A seemingly alternative procedure to the one described above switches the order of Steps 2 and 4, such that the hypothetical assignment vector is drawn first, and the missing compliance statuses are imputed second.  This alternative procedure, however, is exactly equivalent to the one described above because the imputation of the missing compliance statuses under the null hypothesis is influenced by $Z$ only through $C^\mathrm{obs}$.  Because $C^{\mathrm{obs}}$ is fixed by the actual observed data, reversing the order of Steps 2 and 4 does not affect the overall inferential procedure.  Intuitively, we can consider the posterior predictive $p$-value as a double integral over the missing compliance statuses and the randomization; switching the order of integration does not affect the result.

\subsection{Illustrative simulations with non-compliance}\label{sec:simNoncompliance}
Consider this modified example from \citet{Rubin1998}: suppose a completely randomized double-blind experiment is conducted to investigate the effect of a new drug (provided in addition to standard care) versus standard care alone on $Y$, which measures the severity of patients' heart attacks in the year following treatment.  $Y$ is ordinal, taking on values of 0, 1, and 2 (no, mild, and severe attacks, respectively).  We assume that all of the patients survive through the year.  We also assume one-sided non-compliance, so our experiment has two groups of patients: compliers and never-takers.

In our simulation, we randomly select $N=1000$ units from a super-population of 10\% compliers and 90\% never-takers; the compliers tend to be healthier than the never-takers.  We randomly assign $N/2=500$ units to control and $N/2$ units to active treatment, observing only the compliance statuses of units assigned to active treatment.  For each unit, we generate an observed Multinomial outcome, $Y_i^\mathrm{obs}$, according to the specified treatment effect hypothesis.  Simulation details are provided in Appendix \ref{apx:marginal1}.

\begin{figure}[!ht]
\centering
\includegraphics[scale=.53]{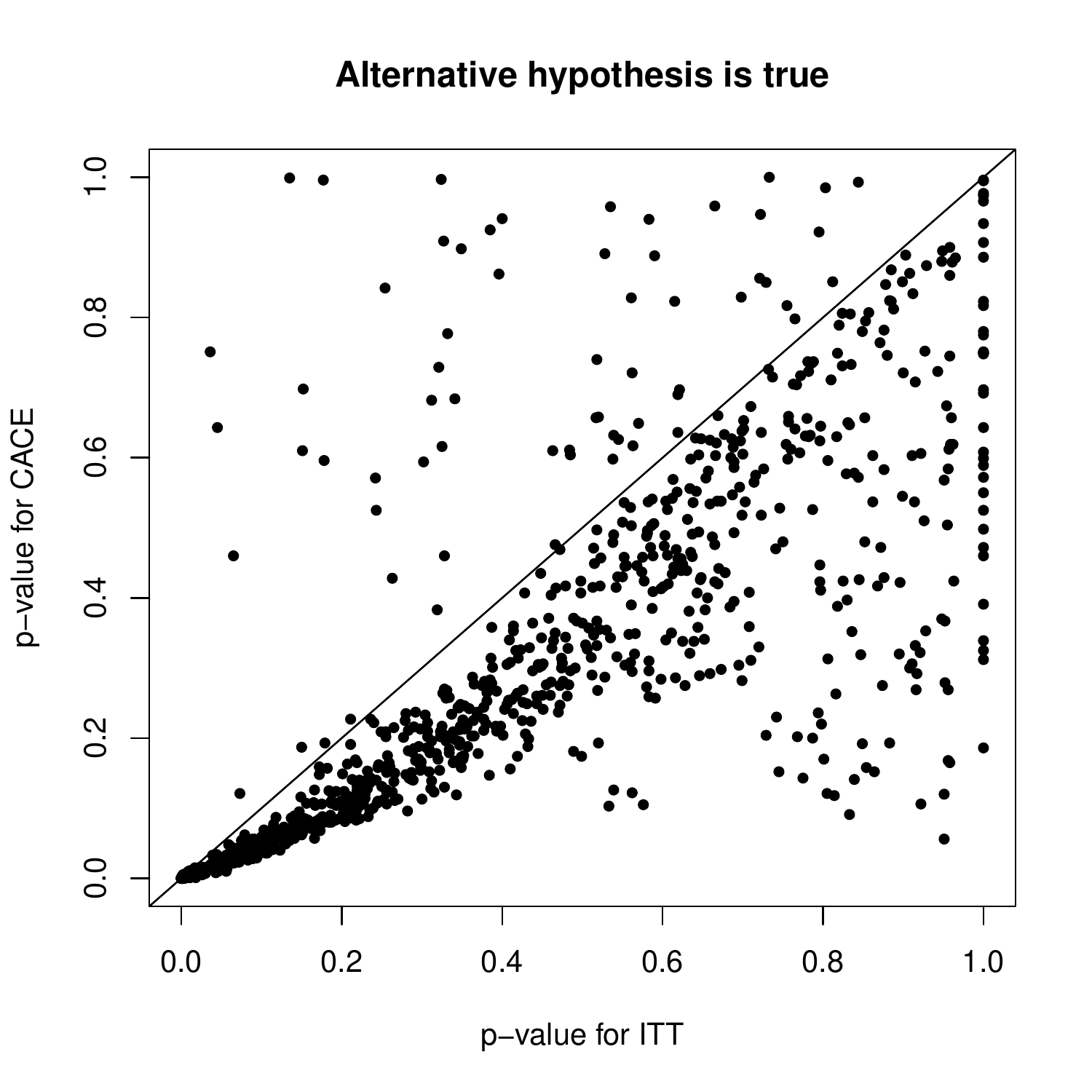}
\caption{Joint distribution of 1,000 posterior predictive $p$-values for ITT and CACE estimates under the alternative hypothesis.  Tests for CACE are more powerful because $p$-values for CACE tend to be lower.}
\label{fig:PvalueComparison}
\end{figure}

Using the simulated observed data, we calculate two test statistics: (i) the ITT statistic, and (ii) the MLE of CACE under the exclusion restriction.  We then calculate randomization-based posterior predictive $p$-values for both test statistics, as described in Section \ref{sec:ppp}, under the null hypothesis of zero treatment effect.  (For the multiple imputation of the missing compliance statuses, we place conjugate Beta$(1,1)$ priors on the parameters governing the complier and never-taker outcome distributions.)  To evaluate the frequency characteristics of the posterior predictive $p$-values, we run 1,000 replications of the data simulation and $p$-value procedures.  Under the null hypothesis, $p$-values for the two statistics both appear valid in terms of type I error; their empirical distributions are approximately uniform.  At the $\alpha=.05$ level, tests on ITT and CACE reject the null hypothesis in 4.5\% and 4.1\% of simulations, respectively.  Under the alternative hypothesis, tests based on the CACE are more powerful (see Figure \ref{fig:PvalueComparison}), with tests on ITT and CACE rejecting the null hypothesis in 16.7\% and 25.2\% of simulations, respectively, at $\alpha=.05$.  In a general setting, the magnitude of the power gain from the CACE depends on the proportion of compliers, the magnitude of the treatment effect, and the $\alpha$ level.

\section{Experiments with Multiple Testing}\label{sec:Multiple}

\subsection{Randomization-based multiple comparisons adjustments}
Suppose we have data from a randomized experiment with $J$ estimands and are interested in testing whether the active treatment has any non-null effects.  The desire for $J$ estimands may result, for example, from multiple outcomes per unit or from multiple, potentially overlapping subgroups of units.  \citet*{Brown1981} and \citet*{Westfall1989} first proposed permutation-based multiple comparisons adjustments, with the latter showing that such adjustments outperform traditional (e.g., Bonferroni) adjustments in terms of power.  They did not, however, explicitly motivate their methods using randomized assignment mechanisms and joint randomization distributions.  Furthermore, they assumed specific models that facilitated the calculation of nominal (unadjusted) $p$-values and implicitly assumed completely randomized assignments throughout.

Here we extend their ideas to a fully randomization-based procedure for multiple comparisons adjustments.  In contrast to the aforementioned work, our procedure is connected to --- and directly motivated by --- the actual randomized assignment mechanism used in the experiment; in addition, both the nominal and adjusted $p$-values in our procedure are randomization-based, so we do not require any assumptions about the underlying distribution of the data.  We calculate fully randomization-based adjusted $p$-values as follows:
\begin{enumerate}
\item \textbf{Choose test statistics and calculate their observed values.} \newline Choose test statistics, $(T_1,\ldots,T_J)$, and calculate $(T_1^\mathrm{obs},\ldots,T_J^\mathrm{obs})$ on the observed data.
\item \textbf{Impute missing potential outcomes.} \newline Impute the missing potential outcomes under the sharp null hypothesis.
\item \textbf{Calculate nominal $p$-values for the observed test statistics.} \newline For $j=1,\ldots,J$, calculate the randomization-based $p$-value for $T_j^\mathrm{obs}$ by repeatedly (i.e., $M'$ times) (i) drawing a random hypothetical assignment vector according to the assignment mechanism, and (ii) calculating the test statistic, $T_j^\mathrm{hyp}$, for the corresponding hypothetical observed data.  The nominal, marginal randomization-based $p$-value for $T_j^\mathrm{obs}$ ($j=1,\ldots,J$) equals the proportion of $T_j^\mathrm{hyp}$ values that are as extreme as or more extreme than $T_j^\mathrm{obs}$.  Record the $(T_1^\mathrm{hyp},\ldots,T_J^\mathrm{hyp})$ values for use in Step 4.
\item \textbf{Calculate nominal (marginal) $p$-values for the hypothetical test statistics.} \newline Using the $M'$ sets of $(T_1^\mathrm{hyp},\ldots,T_J^\mathrm{hyp})$ values from Step 3, calculate a nominal randomization-based $p$-value for each $T_j^\mathrm{hyp}$ and record the minimum of the $p$-values for each of the $M'$ sets.
\item \textbf{Obtain the joint randomization distribution of the nominal $p$-values.} \newline For large $M'$, the repetitions of Step 4 appropriately capture the joint randomization distribution of the test statistics and thus, of the nominal $p$-values.
\item \textbf{Calculate adjusted $p$-values for the observed test statistics.} \newline The adjusted $p$-value \citep*{Westfall1989} for $T_j^\mathrm{obs}$ ($j=1,\ldots,J$) equals the proportion of hypothetical observed datasets for which the minimum of the $J$ nominal $p$-values for $(T_1^\mathrm{hyp},\ldots,T_J^\mathrm{hyp})$ is less than or equal to the nominal $p$-value for $T_j^\mathrm{obs}$.
\end{enumerate}

Steps 4--5 essentially represent a translation, i.e., re-scaling, of hypothetical test statistics --- which may have different scales --- into hypothetical $p$-values, which share a common 0--1 scale.  The procedure described above results in individual adjusted $p$-values that are corrected for the FWER but are also directly interpretable on their own.

Equivalently, to determine $\alpha$-level significance, we can compare each nominal $p$-value to the familywise $\alpha$-level cutoff: the $\alpha$-th quantile of the minimums recorded in Step 4.  The probability that no type I errors are made (i.e., that we fail to reject all $J$ tests under the null hypothesis) is equivalent to the probability that all $J$ observed marginal $p$-values are above the cutoff.  This equals the probability that the minimum of the $J$ observed $p$-values is above the cutoff, which is $1-\alpha$ by construction.  Thus, the probability of at least one type I error --- the FWER --- is $\alpha$, as desired.

Randomization-adjusted $p$-values are more powerful than traditional Bonferroni-adjusted $p$-values, especially when the correlations among the $J$ test statistics are high, as shown by the simulations below.  Intuitively, suppose the null hypothesis of zero effects is true and that we have a large number of uncorrelated test statistics; the probability of at least one type I error is quite high because of the number of tests being conducted.  Now suppose instead that those test statistics are highly correlated; the probability of at least one type I error is reduced because the tests' type I errors are likely to occur simultaneously, i.e., for the same random assignments.  In fact, if the test statistics are perfectly correlated, there is essentially only one test being conducted, so no multiple comparisons adjustment is needed.  Bonferroni adjustments in all of these settings are the same, simply counting the number of $p$-values being examined.  In contrast, by utilizing the joint distribution of the nominal $p$-values, the randomization-based adjustments account for the correlations among test statistics and are less conservative.

\subsection{Illustrative simulations with multiple testing}\label{sec:simMultiple}
We follow the experimental setup from Section \ref{sec:simNoncompliance}, modified to include multiple outcomes but without non-compliance.  Suppose that researchers now want to investigate the effect of the new drug on three outcomes: $Y_{\cdot 1}, Y_{\cdot 2}$, and $Y_{\cdot 3}$ (with the first subscript denoting the participant), which measure the severity of heart attacks (defined as before) in the first, second, and third year after treatment, respectively.  We assume that all of the patients survive through the third year, and we would like to see whether the drug has an effect on heart attack severity at any of the three time points.

To evaluate the frequency characteristics of the adjusted randomization-based $p$-values, we simulate 1,000 datasets under both null and alternative hypotheses according to each of three outcome correlation structures: zero, partial (approximately $0.5$), and perfect correlation.  The specific data generation processes are found in in Appendices \ref{apx:marginal2} and \ref{apx:corr}.  The correlations among $Y_{i1}(z)$, $Y_{i2}(z)$, and $Y_{i3}(z)$ ($z=0,1$) are important; however, for a fixed $j$, the correlation between $Y_{ij}(0)$ and $Y_{ij}(1)$ is inconsequential to the simulation because we only ever observe one of the potential outcomes.

For each simulated dataset, we calculate fully randomization-based adjusted $p$-values and decide whether or not to reject the null hypothesis of zero treatment effects across the three time periods at $\alpha=.05$.  For comparison, we also decide whether or not to reject the null hypothesis using Bonferroni-adjusted $p$-values.  Simulation results under both null and alternative hypotheses are shown in Table \ref{tbl:propreject}.  Without sacrificing validity under the null hypothesis, the randomization-based adjustment displays greater power than the Bonferroni adjustment under the alternative hypothesis, particularly for scenarios with high correlations among outcomes.

\begin{table}[!ht]
\centering
\begin{tabular}{l c c c c}
  \hline
  & \multicolumn{4}{c}{Rejection Rate at $\alpha=.05$} \\ \cline{2-5}
  & \multicolumn{2}{c}{Null is true} & \multicolumn{2}{c}{Alternative is true} \\ \cline{2-5}
  & Bonferroni & Randomization-Based & Bonferroni & Randomization-Based \\ \hline \hline
  Zero correlation & .042 & .046 & .908 & .919 \\
  Partial correlation & .045 & .053 & .787 & .811 \\
  Perfect correlation & .024 & .045 & .557 & .720 \\ \hline
\end{tabular}
\caption{Proportions of multiple testing simulations in which the null hypothesis was rejected, under various data generation processes.  Based on 1,000 replications.}
\label{tbl:propreject}
\end{table}




\section{Experiments with Both Non-compliance and Multiple Testing} \label{sec:combined}
It is natural to merge the analysis methods presented in Sections \ref{sec:Non-compliance} and \ref{sec:Multiple} --- both of which use the randomized assignment mechanism to aid inference --- for experiments involving both non-compliance and multiple testing.  The results are valid familywise tests that are more powerful from both perspectives: more powerful than both those based on standard ITT statistics and those using traditional multiple comparison adjustments.

Suppose again that we have data from a randomized experiment with $J$ estimands and that we are interested in testing whether the active treatment has any non-null effects.  However, not all units comply to their treatment assignments; assume for simplicity that non-compliance is one-sided.  In Section \ref{sec:Non-compliance}, Table \ref{tbl:science} displays the observed values of a Science table with two $Y$ potential outcomes --- one observed and one missing --- for each unit.  Here, Table \ref{tbl:obsScienceMult} shows the corresponding observed values of a Science table with multiple estimands resulting from $J=3$ outcomes of interest.  Each unit has six potential outcomes, only three of which are observed; the other three are missing.  Within unit $i$, we observe the same member of $(Y_{ij}(0),Y_{ij}(1))$ for each outcome $j$, e.g., if we observe $Y_{i1}(1)$, then we also observe $Y_{i2}(1)$ and $Y_{i3}(1)$.

\begin{table}[!ht]
\centering
\resizebox{16cm}{!}{
\begin{tabular}{c c c c c c c c c c c c}
  \hline
  & & Assignment & \multicolumn{2}{c}{$D(z)$} & Compliance status & \multicolumn{2}{c}{$Y_{\cdot 1}(z)$} & \multicolumn{2}{c}{$Y_{\cdot 2}(z)$} & \multicolumn{2}{c}{$Y_{\cdot 3}(z)$} \\ \cline{3-12}
  Unit & $X_i$ & $Z_i$ & $D_i(0)$ & $D_i(1)$ & $C_i$ & $Y_{i1}(0)$ & $Y_{i1}(1)$ & $Y_{i2}(0)$ & $Y_{i2}(1)$ & $Y_{i3}(0)$ & $Y_{i3}(1)$ \\ \hline \hline
  1 & $X_1$ & 0 & 0 & ? & ? & $Y_{11}^{\mathrm{obs}}$ & ? & $Y_{12}^{\mathrm{obs}}$ & ? & $Y_{13}^{\mathrm{obs}}$ & ? \\
  2 & $X_2$ & 1 & 0 & 1 & $c$ & ? & $Y_{21}^{\mathrm{obs}}$ & ? & $Y_{22}^{\mathrm{obs}}$ & ? & $Y_{23}^{\mathrm{obs}}$ \\
  3 & $X_3$ & 1 & 0 & 1 & $c$ & ? & $Y_{31}^{\mathrm{obs}}$ & ? & $Y_{32}^{\mathrm{obs}}$ & ? & $Y_{33}^{\mathrm{obs}}$ \\
  4 & $X_4$ & 1 & 0 & 0 & $nt$ & ? & $Y_{41}^{\mathrm{obs}}$ & ? & $Y_{42}^{\mathrm{obs}}$ & ? & $Y_{43}^{\mathrm{obs}}$ \\
  \ldots & & & \multicolumn{2}{c}{\ldots} & & & & \multicolumn{2}{c}{\ldots} & & \\
  $N$ & $X_N$ & 0 & 0 & ? & ? & $Y_{N1}^{\mathrm{obs}}$ & ? & $Y_{N2}^{\mathrm{obs}}$ & ? &$Y_{N3}^{\mathrm{obs}}$ & ? \\
  \hline
\end{tabular}
}
\caption{Observed and unobserved values of the Science table from Table \ref{tbl:science}, now with three outcomes of interest.  Missing (unobserved) data are denoted by question marks.}
\label{tbl:obsScienceMult}
\end{table}

In experiments with non-compliance and multiple testing, obtaining valid and more powerful familywise tests involves (i) calculating (nominal) posterior predictive $p$-values for CACE according to the procedure in Section \ref{sec:Non-compliance}, and (ii) calculating adjusted posterior predictive $p$-values using the joint randomization distribution of the nominal $p$-values, according to the procedure in Section \ref{sec:Multiple}.  Intuitively, this combined method of analysis is preferable because Steps (i) and (ii) provide power gains through distinct and unrelated mechanisms, and neither sacrifices validity in terms of type I error.  For the $J$ estimands, we expect each individual (nominal) CACE $p$-value to be more powerful than its ITT counterpart based on the arguments in Section \ref{sec:Non-compliance}.  Furthermore, given a set of $J$ nominal $p$-values, we expect randomization-adjusted $p$-values using the nominal $p$-values' joint randomization distribution to be more powerful than Bonferroni-adjusted $p$-values, as argued in Section \ref{sec:Multiple}.  Naturally, adjusting more powerful nominal $p$-values in a more powerful manner results in adjusted $p$-values that are doubly advantageous in terms of power.  The full procedure is detailed below:
\begin{enumerate}
\item \textbf{Choose test statistics and calculate their observed values.} \newline Choose test statistics, $(T_1,\ldots,T_J)$, and calculate $(T_1^\mathrm{obs},\ldots,T_J^\mathrm{obs})$ on the actual observed data.
\end{enumerate}
\begin{algorithmic}
\For{$i:1$ to $M$}
\begin{enumerate} \setcounter{enumi}{1}
\item \textbf{Impute missing compliance statuses.} \newline Impute the missing compliance statuses, drawing once from their posterior predictive distribution according to a compliance model that assumes the null hypothesis.
\item \textbf{Impute missing potential outcomes.} \newline Impute all of the missing $(Y_1,\ldots,Y_J)$ potential outcomes under the sharp null hypothesis.
\item \textbf{Draw a random hypothetical assignment.} \newline Draw a random hypothetical assignment vector according to the assignment mechanism.
\item \textbf{Re-observe the data.} \newline Treating the imputed compliance statuses and potential outcomes and the hypothetical assignment vector as true, create a corresponding hypothetical observed dataset by masking the potential outcomes and compliance statuses that would not have been observed under the hypothetical assignment.
\item \textbf{Calculate test statistics on the hypothetical observed data.} \newline Calculate $(T_1,\ldots,T_J)$ on the hypothetical observed data to obtain $(T_1^\mathrm{hyp},\ldots,T_J^\mathrm{hyp})$.  For $j=1,\ldots,J$, record whether $T_j^\mathrm{hyp}$ is at least as extreme as $T_j^\mathrm{obs}$.
\end{enumerate}
\EndFor
\end{algorithmic}
\begin{enumerate} \setcounter{enumi}{6}
\item \textbf{Calculate nominal (marginal) posterior predictive $p$-values for the observed test statistics.} \newline For $j=1,\ldots,J$, the nominal (marginal) posterior predictive $p$-value for the null hypothesis with respect to the test statistic $T_j$ equals the proportion of the $M$ imputation-randomization sets created by Steps 2--6 for which $T_j^\mathrm{hyp}$ is as extreme as or more extreme than $T_j^\mathrm{obs}$.
\item \textbf{Calculate nominal posterior predictive $p$-values for the hypothetical test statistics and obtain the joint randomization distribution of the nominal posterior predictive $p$-values.} \newline For each of the $M$ imputation-randomization sets, translate the hypothetical test statistics $(T_1^\mathrm{hyp},\ldots,T_J^\mathrm{hyp})$ into hypothetical nominal posterior predictive $p$-values using proportions similar to the one described in Step 7.  This step is a computationally efficient way of obtaining the joing distribution of hypothetical test statistics on a common $p$-value scale, analogous to Steps 4--5 from the procedure in Section \ref{sec:Multiple}.  Record the minimum of each set of nominal $p$-values.
\item \textbf{Calculate adjusted posterior predictive $p$-values for the observed test statistics.} \newline The adjusted posterior predictive $p$-value for $T_j^\mathrm{obs}$ ($j=1,\ldots,J$) equals the proportion of the $M$ imputation-randomization sets for which the minimum of the $J$ nominal posterior predictive $p$-values for $(T_1^\mathrm{hyp},\ldots,T_J^\mathrm{hyp})$ is less than or equal to the nominal (marginal) posterior predictive $p$-value for $T_j^\mathrm{obs}$.
\end{enumerate}

Under the null hypothesis, the outcomes $Y_{\cdot 1},\ldots,Y_{\cdot J}$ inform the multiple imputation of the missing compliance statuses.  Posterior predictive imputations of the missing compliance statuses can be generated using a data augmentation algorithm similar to the one described in Section \ref{sec:Non-compliance}, with Equation \ref{eqn:compprob} modified to use the joint set of $J$ observed outcomes.

\subsection{Illustrative simulations with both non-compliance and multiple testing}
Again consider the heart treatment example from Sections \ref{sec:simNoncompliance} and \ref{sec:simMultiple}: we would like to see whether the active treatment has an effect on heart attack severity at any of the three time points after treatment.  In these simulations, we assume one-sided non-compliance, with $N=1000$ units randomly sampled from super-populations with 10\% and 30\% compliers.  We also ran simulations with 50\% and 70\% compliance rates, but almost all of the tests were able to detect treatment effects under the alternative hypotheses, so the comparison tables were less meaningful.  Alternative hypotheses 1, 2, and 3, in that order, assume treatment effects of increasing size.  The data generation processes are described in Appendices \ref{apx:marginal3} and \ref{apx:corr}.

For each simulated dataset, a total of 10 familywise tests are conducted.  Five of the tests use the ITT test statistic: one uses the Bonferroni correction and one uses the randomization-based multiple comparisons adjustment.  The other three ITT tests use multiple comparisons adjustments proposed as alternatives to the Bonferroni correction, by \citet{Holm1979}, \citet{Hochberg1988}, and \citet{Hommel1988}.  The remaining five tests use the MLE of CACE (under the exclusion restriction) as the test statistic instead of the ITT test statistic.  Table \ref{tbl:propreject2} displays proportions of simulations in which the null hypothesis was rejected, based on 1000 replications.

\begin{table}[!ht]
\centering
\resizebox{16cm}{!}{
\begin{tabular}{l c c c c c | c c c c c}
  \hline
  Compliance Rate = 0.1 & \multicolumn{10}{c}{Rejection Rate at $\alpha=.05$} \\ \cline{2-11}
  & \multicolumn{5}{c}{ITT} & \multicolumn{5}{c}{CACE} \\ \cline{2-11}
  Null is true & Bonferroni & Holm & Hochberg & Hommel & Rand-Based & Bonferroni & Holm & Hochberg & Hommel & Rand-Based \\ \hline \hline
  Zero correlation & .035 & .035 & .037 & .039 & .050 & .033 & .033 & .033 & .034 & .039 \\
  Partial correlation & .025 & .025 & .026 & .030 & .041 & .025 & .025 & .026 & .031 & .039 \\
  Perfect correlation & .012 & .012 & .047 & .047 & .049 & .008 & .008 & .032 & .032 & .033 \\ \hline
\end{tabular}
}
\resizebox{16cm}{!}{
\begin{tabular}{l c c c c c | c c c c c}
  \hline
  & \multicolumn{5}{c}{ITT} & \multicolumn{5}{c}{CACE} \\ \cline{2-11}
  Alternative 1 is true & Bonferroni & Holm & Hochberg & Hommel & Rand-Based & Bonferroni & Holm & Hochberg & Hommel & Rand-Based \\ \hline \hline
  Zero correlation & .161 & .161 & .167 & .169 & .189 & .209 & .209 & .217 & .226 & .228 \\
  Partial correlation & .113 & .113 & .118 & .123 & .154 & .179 & .179 & .189 & .196 & .228 \\
  Perfect correlation & .062 & .062 & .139 & .139 & .139 & .094 & .094 & .205 & .205 & .207\\ \hline
\end{tabular}
}
\resizebox{16cm}{!}{
\begin{tabular}{l c c c c c | c c c c c}
  \hline
  & \multicolumn{5}{c}{ITT} & \multicolumn{5}{c}{CACE} \\ \cline{2-11}
  Alternative 2 is true & Bonferroni & Holm & Hochberg & Hommel & Rand-Based & Bonferroni & Holm & Hochberg & Hommel & Rand-Based \\ \hline \hline
  Zero correlation & .303 & .303 & .306 & .310 & .342 & .357 & .357 & .366 & .369 & .380 \\
  Partial correlation & .204 & .204 & .219 & .225 & .273 & .303 & .303 & .314 & .320 & .357 \\
  Perfect correlation & .137 & .137 & .270 & .270 & .270 & .184 & .184 & .364 & .364 & .369 \\ \hline
\end{tabular}
}
\resizebox{16cm}{!}{
\begin{tabular}{l c c c c c | c c c c c}
  \hline
  & \multicolumn{5}{c}{ITT} & \multicolumn{5}{c}{CACE} \\ \cline{2-11}
  Alternative 3 is true & Bonferroni & Holm & Hochberg & Hommel & Rand-Based & Bonferroni & Holm & Hochberg & Hommel & Rand-Based \\ \hline \hline
  Zero correlation & .688 & .688 & .702 & .716 & .754 & .710 & .710 & .746 & .756 & .742 \\
  Partial correlation & .370 & .370 & .384 & .398 & .444 & .474 & .474 & .487 & .499 & .518 \\
  Perfect correlation & .297 & .297 & .465 & .465 & .471 & .357 & .357 & .565 & .565 & .570 \\ \hline
\end{tabular}
}
\newline
\vspace{8mm}
\newline
\resizebox{16cm}{!}{
\begin{tabular}{l c c c c c | c c c c c}
  \hline
  Compliance Rate = 0.3 & \multicolumn{10}{c}{Rejection Rate at $\alpha=.05$} \\ \cline{2-11}
  & \multicolumn{5}{c}{ITT} & \multicolumn{5}{c}{CACE} \\ \cline{2-11}
  Null is true & Bonferroni & Holm & Hochberg & Hommel & Rand-Based & Bonferroni & Holm & Hochberg & Hommel & Rand-Based \\ \hline \hline
  Zero correlation & .037 &.037 & .037 & .038 & .046 & .031 & .031 & .032 & .033 & .038 \\
  Partial correlation & .033 & .033 & .035 & .036 & .044 & .025 & .025 & .025 & .026 & .035 \\
  Perfect correlation & .010 & .010 & .032 & .032 & .035 & .007 & .007 & .038 & .038 & .039 \\ \hline
\end{tabular}
}
\resizebox{16cm}{!}{
\begin{tabular}{l c c c c c | c c c c c}
  \hline
  & \multicolumn{5}{c}{ITT} & \multicolumn{5}{c}{CACE} \\ \cline{2-11}
  Alternative 1 is true & Bonferroni & Holm & Hochberg & Hommel & Rand-Based & Bonferroni & Holm & Hochberg & Hommel & Rand-Based \\ \hline \hline
  Zero correlation & .529 & .529 & .549 & .557 & .595 & .617 & .617 & .632 & .642 & .670 \\
  Partial correlation & .482 & .482 & .511 & .532 & .571 & .604 & .604 & .625 & .638 & .671 \\
  Perfect correlation & .309 & .309 & .492 & .492 & .497 & .420 & .420 & .606 & .606 & .611 \\ \hline
\end{tabular}
}
\resizebox{16cm}{!}{
\begin{tabular}{l c c c c c | c c c c c}
  \hline
  & \multicolumn{5}{c}{ITT} & \multicolumn{5}{c}{CACE} \\ \cline{2-11}
  Alternative 2 is true & Bonferroni & Holm & Hochberg & Hommel & Rand-Based & Bonferroni & Holm & Hochberg & Hommel & Rand-Based \\ \hline \hline
  Zero correlation & .907 & .907 & .914 & .914 & .923 & .969 & .969 & .978 & .979 & .981 \\
  Partial correlation & .859 & .859 & .868 & .872 & .891 & .946 & .946 & .960 & .963 & .968 \\
  Perfect correlation & .752 & .752 & .862 & .862 & .865 & .871 & .871 & .957 & .957 & .957 \\ \hline
\end{tabular}
}
\resizebox{16cm}{!}{
\begin{tabular}{l c c c c c | c c c c c}
  \hline
  & \multicolumn{5}{c}{ITT} & \multicolumn{5}{c}{CACE} \\ \cline{2-11}
  Alternative 3 is true & Bonferroni & Holm & Hochberg & Hommel & Rand-Based & Bonferroni & Holm & Hochberg & Hommel & Rand-Based \\ \hline \hline
  Zero correlation & .993 & .993 & .995 & .996 & .996 & .999 & .999 & .999 & .999 & .999 \\
  Partial correlation & .984 & .984 & .986 & .987 & .990 & .997 & .997 & .998 & .999 & .999 \\
  Perfect correlation & .966 & .966 & .993 & .993 & .993 & .989 & .989 & .999 & .999 &.999 \\ \hline
\end{tabular}
}
\caption{Proportions of simulations in which the null hypothesis was rejected, under various data generation processes.  Based on 1000 replications.}
\label{tbl:propreject2}
\end{table}

Under the null hypotheses, all 10 familywise tests appear valid in terms of type I error.  The randomization-based tests have the rejection rates closest to the nominal rejection rates.  As expected, the Bonferroni-adjusted tests are conservative, especially when correlation among outcomes is high.  In such settings, there are, in a sense, fewer possible effects to detect, and randomization-adjusted rejection rates are much higher relative to their Bonferroni-adjusted counterparts.  The Holm, Hochberg, and Hommel procedures all perform similarly to Bonferroni under the null hypotheses.


Under alternative hypotheses, the CACE tests generally have higher power, i.e., higher rejection rates, than the ITT tests.  In addition, the randomization-based tests perform very well, especially when correlation among outcomes is high.  The Bonferroni and Holm tests perform extremely similarly, while the Hochberg and Hommel tests perform slightly better.  The randomization-based procedure generally outperforms all four of the other procedures.  In our simulations, CACE tests with randomization-based multiple comparisons adjustments have up to 3.3 times the power of traditional Bonferroni ITT tests, when treatment effects are difficult to detect.  The relative power gain is less pronounced when treatment effects are larger, though gains are still apparent in the absolute scale.  In a particular experimental setting, the magnitude of the power gain from the combined analysis method depends on the compliance rate, the magnitude of the treatment effect, the $\alpha$ level, and the correlation of the multiple test statistics.

\section{The National Job Training Partnership Act Study}\label{sec:JTPA}
Title II of the United States Job Training Partnership Act (JTPA) of 1982 funded employment training programs for economically disadvantaged residents \citep*{Bloom1997,Abadie2002}.  To evaluate the effectiveness of those training programs, the National JTPA Study conducted a randomized experiment through 16 local administration areas involving a total of around 20,000 participants who applied for JTPA services from November 1987 to September 1989 \citep*{Upjohn2013}.  Treatment group participants were eligible to receive JTPA services, while control group participants were ineligible to receive JTPA services for 18 months.  Not every participant assigned to the treatment group actually enrolled and received JTPA services.

\subsection{The data}
Monthly employment outcomes were recorded for 30 months after assignment through follow-up surveys and administrative records from state unemployment insurance agencies.  Researchers were interested in measuring JTPA effects across three time periods representing various stages of training and employment: months 1--6 (after assignment), the period when most JTPA enrollees were in the program; months 7--18, approximately the first post-program year; and months 19--30, approximately the second post-program year \citep*{Bloom1997}.

\citet*{Bloom1997}'s original JTPA report evaluates effects on average income but does not explicitly address the large portion of zero-income (i.e., unemployed) participants.  Although the report describes effects by subperiod as well as by various participant subgroups, it fails to mention or employ any multiple comparisons adjustments.  Here we focus on JTPA's effects on employment status and use gender as our only background covariate; this facilitates standard, non-controversial modeling choices (see Section \ref{sec:JTPAmodel}) and allows us to highlight our methodological contributions rather than discuss the sensitivity of our results to various, possibly complicated modeling decisions.  Our methods can be extended to evaluate effects on other outcome variables, such as income and wages, provided that we outline a reasonable imputation model \citep*{Zhang2009}.

We would like to evaluate whether JTPA had an effect on employment status for any of the three time periods.  Because employment characteristics often differ by gender, we examine JTPA effects for the three time periods by gender, for a total of six gender-time groups.  For illustrative purposes, we restrict our study population to adults who had obtained a high school or GED diploma (7,445, or 66.4\%, of the 11,204 total adults in the original JTPA study) and assume complete randomization (with an approximate $2:1$ treatment-to-control assignment ratio) of the participants, ignoring the local administration structure because of the limitations of the available data.

Of the 5,009 participants assigned to the treatment group, 3,316 (66.2\%) subsequently received JTPA training.  Although the study protocol barred participants assigned to the control group from receiving JTPA services for 18 months, 41 (1.7\%) of 2,436 adults in the control group did in fact receive services within that time frame.  To create a simpler setting with true one-sided non-compliance, we discard these 41 participants (0.6\% of the 7,445 total adults in our study) with the belief that their inclusion would have a negligible influence on the resulting inference.

\begin{figure}[!ht]
\centering
\includegraphics[scale=.45]{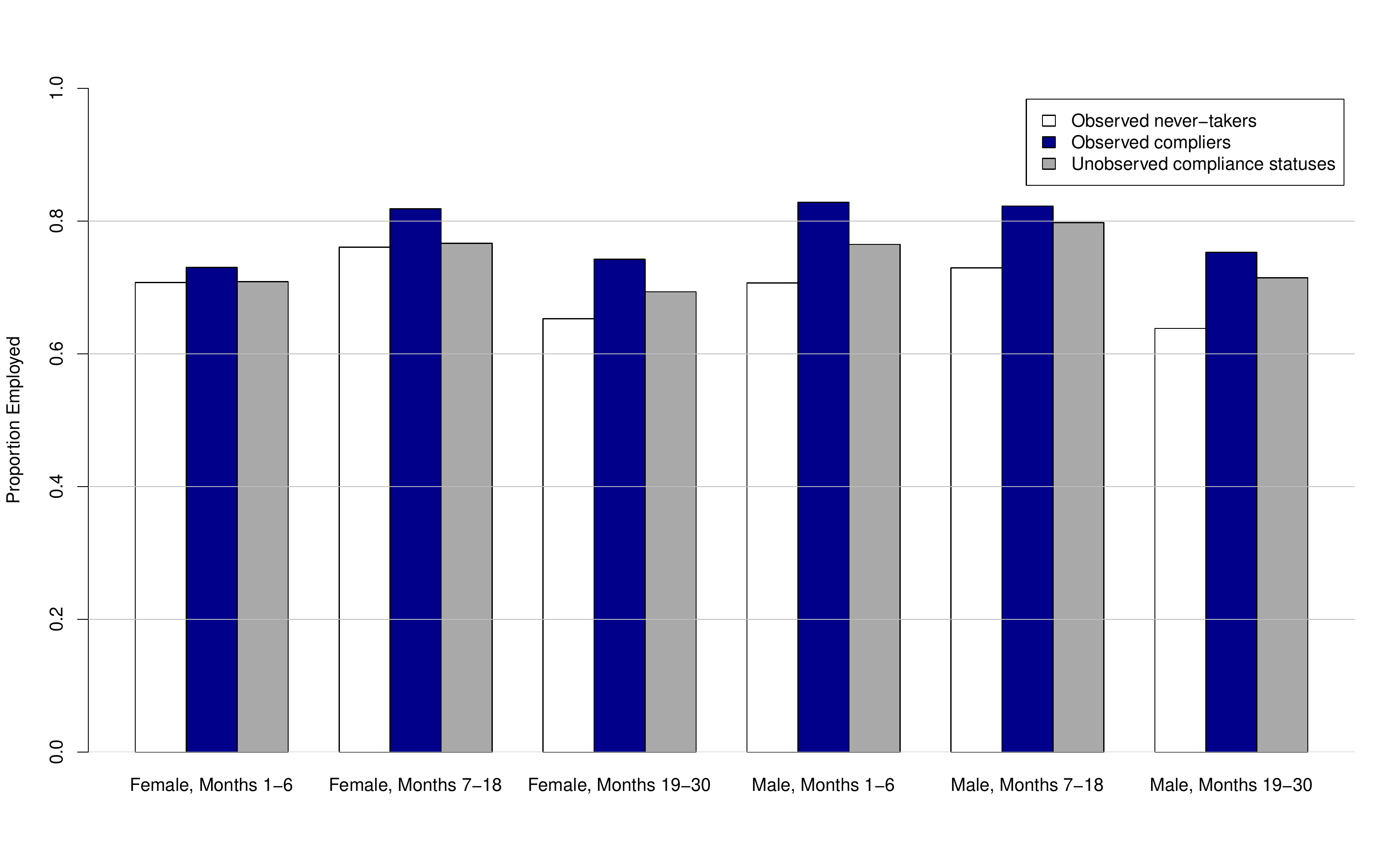}
\caption{Observed employment proportions for JTPA participants by compliance status across the six gender-time groups.}
\label{fig:observedProportions}
\end{figure}

Given two genders and three time periods, we have six complier-focused estimands in total, each one representing the difference in employment proportions within a particular gender-time group when receiving versus not receiving JTPA services.  Two summaries of the observed data are provided in Figure \ref{fig:observedProportions} and Table \ref{tbl:observedCompliance}.  Figure \ref{fig:observedProportions} shows observed employment proportions across the six gender-time groups by observed compliance status.  Within every group, observed compliers are employed at a higher rate than observed never-takers.  Participants with unobserved compliance statuses (i.e., those assigned to control) are a mixture of compliers and never-takers, and tend to be employed at a rate in between the rates for observed compliers and observed never-takers.

\begin{table}[!ht]
\centering
\begin{tabular}{l c c}
  \hline
  & \multicolumn{2}{c}{Observed Employment Proportions} \\ \cline{2-3}
  & Assigned Control & Assigned Treatment \\
  & $C_i \in \{ c,nt \}, Z_i=0$ & $C_i \in \{ c,nt \}, Z_i=1$ \\ \hline \hline
  Female, Months 1--6 & .709 & .723 \\
  Female, Months 7--18 & .767 & .800 \\
  Female, Months 19--30 & .694 & .714 \\
  Male, Months 1--6 & .765 & .785 \\
  Male, Months 7--18 & .798 & .789 \\
  Male, Months 19--30 & .715 & .712 \\ \hline
  & Received Control & Received Treatment \\
  & $(C_i \in \{ c,nt \}, Z_i=0$) or ($C_i=nt, Z_i=1$) & $C_i=c, Z_i=1$ \\ \hline \hline
  Female, Months 1--6 & .708 & .730 \\
  Female, Months 7--18 & .764 & .818 \\
  Female, Months 19--30 & .677 & .743 \\
  Male, Months 1--6 & .740 & .828 \\
  Male, Months 7--18 & .769 & .823 \\
  Male, Months 19--30 & .683 & .753 \\ \hline
\end{tabular}
\caption{Observed employment proportions across the six gender-time groups according to both assignment to and receipt of JTPA services.}
\label{tbl:observedCompliance}
\end{table}

Table \ref{tbl:observedCompliance} displays observed employment proportions across the gender-time groups according to both treatment assignment and treatment receipt, with the corresponding compliance compositions.  We see that participants who received JTPA services, all of whom are compliers, tend to be employed at a higher rate than participants who were merely assigned to the treatment group (a mixture of compliers and never-takers), corroborating the findings in Figure \ref{fig:observedProportions} and suggesting that CACE statistics may lead to more significant estimated effects.  In addition, we observe that participants who did not receive JTPA services --- including any participants assigned to control as well as the never-takers assigned to JTPA --- are employed at a lower rate than just the participants assigned to control.  This inequality is intuitive because the observed never-takers are shown in Figure \ref{fig:observedProportions} to be employed at a lower rate than the assigned control group.

\subsection{Imputation model for CACE}\label{sec:JTPAmodel}
To test the null hypothesis of zero effects using the CACE statistic specified in Section \ref{sec:Non-compliance}, we must specify an imputation model for the missing compliance statuses.  Let $X_i$ and $\bm{Y_i}$ denote the gender and the length-3 vector of employment outcomes (across the three time periods) of participant $i$.  The three elements of $\bm{Y_i}$ are binary, so there are $2^3=8$ possible values of $\bm{Y_i}$; we model $\bm{Y}$ as a Multinomial random variable with eight categories.  Let $\omega_c$ be the super-population proportion of compliers, and let $\bm{\eta}=(\eta_{fc},\eta_{fn},\eta_{mc},\eta_{mn})$ be the parameters that govern the outcome distributions of female compliers, female never-takers, male compliers, and male never-takers, respectively.  Under the null hypothesis, these are the only four outcome distributions because we disregard treatment assignment.  We place a conjugate Beta(1,1) prior on $\omega_c$ and independent conjugate Dirichlet($\bm{1}$) priors on the four $\eta$ parameters, where $\bm{1}$ is a length-8 vector of 1's.

Conditional on $\bm{\eta}$ and a participant's gender and compliance status, the natural outcome distribution under the null hypothesis is: $$\bm{Y_i}^\mathrm{obs}|X_i=x,C_i=q,\bm{\eta} \sim \mbox{Multinomial}(1,\eta_{xq}).$$  Note that we do not assume that the three employment outcomes are independent; this model is fully non-parametric for the joint distribution of the three outcomes.
The posterior distributions of $\omega_c$ and $\bm{\eta}$ are informed by the outcomes of the participants with observed compliance statuses, i.e., those assigned to active treatment, and remain Beta and Dirichlet, respectively.  For each gender $x$ and compliance status $q$, write the Multinomial probability vector as
$$\eta_{xq}=(\pi_{xq1},...,\pi_{xq7},1-\pi_{xq1}-\ldots-\pi_{xq7}).$$
Let $$g_{xq}(\bm{y}; \eta_{xq}) = \pi_{xq1}^{I\{ \bm{y}=(0,0,0) \}} \pi_{xq2}^{I\{ \bm{y}=(0,0,1) \}} \ldots (1-\pi_{xq1}-\ldots-\pi_{xq7})^{I\{ \bm{y}=(1,1,1) \}}$$ denote the probability of outcome $\bm{y}$ for participants of gender $x$ and compliance status $q$.  Then, given a posterior draw of $(\omega_c,\bm{\eta})$, the missing compliance statuses are imputed probabilistically according to Bayes' rule:
\begin{equation}
P(C_i=c|\bm{Y_i^{\mathrm{obs}}},X_i=x,Z_i=0,\omega_c,\bm{\eta})=
\frac{\omega_c g_{xc}(\bm{Y_i^\mathrm{obs}}; \eta_{xc})}{\omega_c g_{xc}(\bm{Y_i^\mathrm{obs}}; \eta_{xc})+(1-\omega_c) g_{xn}(\bm{Y_i^\mathrm{obs}}; \eta_{xn})}.
\end{equation}

\subsection{Results and analysis}
The observed values of the ITT and CACE statistics --- i.e., the estimated effects of JTPA assignment and of receipt, respectively --- are shown in the second column of Table \ref{tbl:jtpa}.  As we expect, the observed CACE values have larger magnitudes; the estimated ITT effects are diluted toward zero by the never-takers, who do not receive any treatment benefit.  Because $\mbox{ITT}=\omega_c*\mbox{CACE}+(1-\omega_c)*0$, the estimated ITT effects are diluted by a proportion equal to one minus the compliance rate.  Due to the random treatment assignment, we expect the overall compliance rate to be approximately equal to the compliance rate observed in the treatment group (66.2\%).

\begin{table}[!ht]
\centering
\begin{tabular}{l c c c c}
  \hline
  & & & \multicolumn{2}{c}{Adjusted $p$-values} \\ \cline{4-5}
  ITT & Estimated Effect & Nominal $p$-value & Bonferroni & Randomization \\ \hline \hline
  Female, Months 1--6 & .014 & .351 & 1.000 & .895 \\
  Female, Months 7--18 & .020 & .199 & 1.000 & .685 \\
  Female, Months 19--30 & .033 & .014 & .085 & .077 \\
  Male, Months 1--6 & -.008 & .582 & 1.000 & .991 \\
  Male, Months 7--18 & .020 & .175 & 1.000 & .636 \\
  Male, Months 19--30 & -.003 & .874 & 1.000 & 1.000 \\ \hline
\end{tabular}
\begin{tabular}{l c c c c}
  & & & \multicolumn{2}{c}{Adjusted $p$-values} \\ \cline{4-5}
  CACE & Estimated Effect & Nominal $p$-value & Bonferroni & Randomization \\ \hline \hline
  Female, Months 1--6 & .020 & .130 & .778 & .302 \\
  Female, Months 7--18 & .034 & .009 & .055 & .026 \\
  Female, Months 19--30 & .049 & .0002 & .001 & .001 \\
  Male, Months 1--6 & -.010 & .462 & 1.000 & .804 \\
  Male, Months 7--18 & .028 & .028 & .169 & .076 \\
  Male, Months 19--30 & -.001 & .967 & 1.000 & 1.000 \\ \hline
\end{tabular}
\caption{Observed values, nominal $p$-values, and Bonferroni- and randomization-adjusted $p$-values for the six JTPA gender-time groups.  Nominal $p$-values are obtained through randomization tests using 10,000 randomizations.}
\label{tbl:jtpa}
\end{table}

Using randomization tests and the methods described in Section \ref{sec:combined}, we obtain one set of nominal ITT $p$-values and a second set of nominal CACE $p$-values, listed in the third column of Table \ref{tbl:jtpa}.  Each set contains six $p$-values, one for each gender-time group.  We also apply Bonferroni and randomization adjustments to both sets of nominal $p$-values, resulting in four total sets of adjusted $p$-values, listed in the rightmost columns of Table \ref{tbl:jtpa}.

The nominal ITT $p$-value for the ``Female, Months 19--30" group indicates statistical significance at the $\alpha=.05$ level.  However, after adjusting for multiple comparisons, neither the Bonferroni- nor randomization-adjusted ITT $p$-values for this group meets the $.05$ threshold.  Across the six gender-time groups, the randomization-adjusted $p$-values tend to be smaller than their Bonferroni-adjusted counterparts; the adjusted $p$-values are tempered less when controlling the FWER via the statistics' joint randomization distribution because of the correlations among the six nominal $p$-values.

Overall, the CACE $p$-values are smaller --- more sensitive to complier-only effects --- than the ITT $p$-values.  In particular, the CACE $p$-values for the ``Female, Months 7--18" and ``Female, Months 19--30" groups indicate a much greater level of significance for the estimated effects of JTPA on employment.  Applying randomization-based instead of Bonferroni adjustments to the CACE $p$-values further increases the indicated significance of these estimated effects.  The small randomization-adjusted CACE $p$-values for these groups suggest that either an event has occurred that is \textit{a priori} rare under the sharp null hypothesis of zero effects, or the sharp null hypothesis is not true --- \textit{receipt} of JTPA services did have an effect on the employment statuses of females with high school or GED diplomas in their first and second post-program years.  The corresponding ITT $p$-values, although smallest among the six groups, are larger and do not have sufficient power to detect an effect on employment status for any of the gender-time groups.

This increase in power is general.  We observe similar $p$-value trends when comparing our methods to ITT and Bonferroni analyses on JTPA data without the high school/GED diploma restriction as well as on other JTPA subgroups analyzed in \citet*{Bloom1997}.


\section{Conclusion} \label{sec:conclusion}
We have detailed a randomization-based procedure for analyzing experimental data in the presence of both non-compliance and multiple testing that is more powerful than traditional ITT and Bonferroni analyses.  As shown through simulations and analyses of the National JTPA Study data, a combined randomization-based procedure can be doubly advantageous, offering gains in power from both perspectives.

The ITT tests for the JTPA Study suggest that the training program had no real effects in increasing employment for either gender at any time point.  The Bonferroni-adjusted CACE tests suggest that JTPA only increased employment for females in the long term (months 19--30).  From a policy perspective, this initiative may be deemed too costly based on the time delay, as well as the fact that all five other subgroups had insignificant effects.  Once we look at the randomization-adjusted CACE tests though, we conclude that JTPA actually had a positive effect on employment for females as soon as they finish the training program, and that the effect sustained into the longer term.  Thus, it seems reasonable for policymakers to fund similar job training programs targeted for women.

\citet*{Westfall1989} assumed Binomial data that facilitate closed-form calculations of nominal $p$-values, which were then adjusted using a permutation test.  Here we propose fully randomization-based $p$-values –-- we exploit the randomization test to calculate \textit{both} nominal and adjusted $p$-values.  In addition, \citet*{Westfall1989} described the adjusted $p$-values as ``permutation-style," not explicitly motivated by the assignment mechanism in a randomized experiment.  In its exploration of non-compliance, \citet*{Rubin1998} required the randomization test to follow the randomized assignment mechanism actually used in the original experiment, an approach we advocate.

A number of other multiple comparisons procedures aim to address the false discovery rate (FDR) \citep*{Benjamini1995}, rather than the FWER.  These two error metrics are conceptually different; the choice of metric should be decided by the researcher depending on the field and specific research setting and goals.  FDR is often preferred in settings with a large number of tests, such as genetic studies, in which finding one true genetic link may outweigh finding a few spurious links.  In such cases, attempting to make exactly zero type I errors can be quite restrictive.  On the other hand, FWER is often used in social science and pharmaceutical settings, in which governmental and regulatory agencies place the onus on the researcher to show that the treatment provides a beneficial effect.  In these cases, the number of tests tends to be smaller, and type I errors can be extremely costly in terms of dollars to taxpayers and health risks (e.g., side effects) to patients.  For these reasons, we focus our discussion on the FWER.

\section*{Acknowledgements}
The first author is supported by the U.S. Department of Defense through the National Defense Science \& Engineering Graduate Fellowship (NDSEG) Program.  We also thank Don Rubin for the interesting and thought-provoking conversations we have had about this work.

\appendix \appendixpage
\section{Marginal Distributions for Simulations}\label{apx:marginals}
\subsection{Non-compliance}\label{apx:marginal1}
For unit $i=1,\ldots,N$, the control potential outcomes for compliers and never-takers have the following marginal distributions:
\begin{equation}\label{eqn:c0}
Y_i(0)|C_i=c \sim \mbox{Multinomial}(.45,.45,.10);
\end{equation}
\begin{equation}\label{eqn:nt}
Y_i(0)|C_i=nt \sim \mbox{Multinomial}(.02,.02,.96).
\end{equation}
Under the null hypothesis, $Y_i(1)$ has the same marginal distribution as $Y_i(0)$ regardless of compliance status.  Under the alternative hypothesis, the complier treatment potential outcomes follow:
\begin{equation}\label{eqn:c1}
Y_i(1)|C_i=c \sim \mbox{Multinomial}(.80,.10,.10),
\end{equation}
while the never-taker treatment potential outcomes follow Equation \ref{eqn:nt}.

\subsection{Multiple testing}\label{apx:marginal2}
For unit $i=1,\ldots,N$ and outcome $j=1,2,3$, the control potential outcomes marginally follow:
\begin{equation}\label{eqn:nullmarginal}
Y_{ij}(0) \sim \mbox{Multinomial}(.45,.45,.10).
\end{equation}
Under the null hypothesis, $Y_{ij}(1)$ has the same marginal distribution as $Y_{ij}(0)$.  Under the alternative hypotheses, the treatment potential outcomes have the following marginal distribution:
\begin{equation}\label{eqn:altmarginal1}
Y_{ij}(1) \sim \mbox{Multinomial}(.50,.45,.05).
\end{equation}

\subsection{Non-compliance and multiple testing}\label{apx:marginal3}
Under the null hypothesis, the potential outcomes follow the marginal distributions described by Equations \ref{eqn:c0} and \ref{eqn:nt} in Appendix \ref{apx:marginal1}.  $Y_i(1)$ has the same marginal distribution as $Y_i(0)$ regardless of compliance status.

Under alternative hypothesis 1, the complier potential outcomes marginally follow:
\begin{equation}
Y_i(0)|C_i=c \sim \mbox{Multinomial}(.45,.45,.10); \qquad Y_i(1)|C_i=c \sim \mbox{Multinomial}(.80,.10,.10)
\end{equation}

Under alternative hypothesis 2, the complier potential outcomes marginally follow:
\begin{equation}
Y_i(0)|C_i=c \sim \mbox{Multinomial}(.30,.60,.10); \qquad Y_i(1)|C_i=c \sim \mbox{Multinomial}(.80,.10,.10)
\end{equation}

Under alternative hypothesis 3, the complier potential outcomes marginally follow:
\begin{equation}
Y_i(0)|C_i=c \sim \mbox{Multinomial}(.25,.55,.20); \qquad Y_i(1)|C_i=c \sim \mbox{Multinomial}(.80,.10,.10)
\end{equation}

\section{Correlation Structure Generation}\label{apx:corr}
To simulate correlation structures among multiple outcomes, we use the following processes utilizing the marginal distributions described in Appendix \ref{apx:marginals}.  For units $i=1,\ldots,N$ and treatment assignment $z=0,1$,
\begin{itemize}
\item Zero correlation: all $Y_{ij}(z)$ ($j=1,2,3$) are drawn independently according to their marginal distributions.
\item Partial correlation: $Y_{i1}(z)$ is drawn according to its marginal distribution.  With probability $1/2$, $Y_{i2}(z)$ is set equal to the drawn value of $Y_{i1}(z)$; otherwise, $Y_{i2}(z)$ is drawn independently according to its marginal distribution.  $Y_{i3}(z)$ is set equal to $Y_{i1}(z)$ with probability $1/3$, set equal to $Y_{i2}(z)$ with probability $1/3$, or drawn independently according to its marginal distribution.
\item Perfect correlation: $Y_{i1}(z)$ is drawn according to its marginal distribution.  Then both $Y_{i2}(z)$ and $Y_{i3}(z)$ are set equal to the drawn value of $Y_{i1}(z)$.
\end{itemize}

\bibliographystyle{asa}
\bibliography{MultNoncompliance}

\end{document}